\documentclass[twocolumn,prb,showpacs]{revtex4}
\usepackage{amsmath,amsthm}
\usepackage{epsfig}
\usepackage{graphics}
\usepackage{graphicx}
\usepackage{dcolumn}
\usepackage{bm}
\usepackage{sidecap}

\begin{document}

\title{Infrared conductivity of hole accumulation and depletion layers in (Ga,Mn)As- and (Ga,Be)As-based electric field-effect devices}
\author{B. C. Chapler,$^{1,\ast}$ S. Mack,$^2$ L. Ju,$^3$ T. W. Elson,$^1$ B. W. Boudouris,$^4$ E. Namdas,$^5$ J. D. Yuen,$^5$ A. J. Heeger,$^5$ N. Samarth,$^6$ M. Di Ventra,$^1$ R. A. Segalman,$^4$ D. D. Awschalom,$^2$ F. Wang,$^3$ and D. N. Basov$^1$}

\affiliation
{$^{1}$Physics Department, University of California-San Diego, La Jolla, California 92093, USA \\
$^{2}$Center for Spintronics and Quantum Computation, University of California-Santa Barbara, California 93106, USA \\
$^{3}$Department of Physics, University of California at Berkeley, Berkeley, California 94720, USA \\
$^{4}$Department of Chemical and Bimolecular Engineering, University of California at Berkeley, Berkeley, California 94720, USA \\
$^{5}$Center for Polymers and Organic Solids, University of California-Santa Barbara, California 93106, USA \\
$^{6}$Department of Physics, The Pennsylvania State University, University Park, Pennsylvania 16802, USA
}
\date{\today}

\begin{abstract} 

We have fabricated electric double-layer field-effect devices to electrostatically dope our active materials, either $x$=0.015 Ga$_{1-x}$Mn$_x$As or $x$=3.2$\times$10$^{-4}$ Ga$_{1-x}$Be$_x$As. The devices are tailored for interrogation of electric field induced changes to the frequency dependent conductivity in the accumulation or depletions layers of the active material via infrared (IR) spectroscopy. The spectra of the (Ga,Be)As-based device reveal electric field induced changes to the IR conductivity consistent with an enhancement or reduction of the Drude response in the accumulation and depletion polarities, respectively. The spectroscopic features of this device are all indicative of metallic conduction within the GaAs host valence band (VB). For the (Ga,Mn)As-based device, the spectra show enhancement of the far-IR itinerant carrier response and broad mid-IR resonance upon hole accumulation, with a decrease of these features in the depletion polarity. These later spectral features demonstrate that conduction in ferromagnetic (FM) Ga$_{1-x}$Mn$_x$As is distinct from genuine metallic behavior due to extended states in the host VB. Furthermore, these data support the notion that a Mn-induced impurity band plays a vital role in the electron dynamics of FM Ga$_{1-x}$Mn$_x$As. We add, a sum-rule analysis of the spectra of our devices suggests that the Mn or Be doping does not lead to a substantial renormalization of the GaAs host VB.

\end{abstract}


\maketitle

The electric field-effect offers a decided advantage in investigations of electronic phenomena in complex materials. Namely, the charge density may be tuned without adding lattice disorder, and while keeping other sample properties intact. In this manner, electrostatic ``doping'' offers a ``clean'' approach to tune insulator-to-metal transitions (IMTs), as well as other phase transitions and electronic behavior~\cite{Ahn2003, Ahn2006, Ueno2008, Qazilbash2008, Chakhalian2012}. In the vast majority of field-effect studies of complex materials, the experiments are limited to transport data. Experiments in the infrared (IR) regime, however, are uniquely suited to probe the evolution of electronic behavior in materials of interest as the charge density is electrostatically tuned. IR experiments serve as a contactless probe, sensitive to the narrow layer of charge accumulation or depletion formed at the interface between the active material and gate insulator~\cite{Allen1975}. Moreover, experiments in the IR regime are favorable because many characteristic energy scales in condensed matter systems fall within the IR range~\cite{Burch2008, Basov2011}. 

\begin{figure}
\centering
\includegraphics[width=86mm]{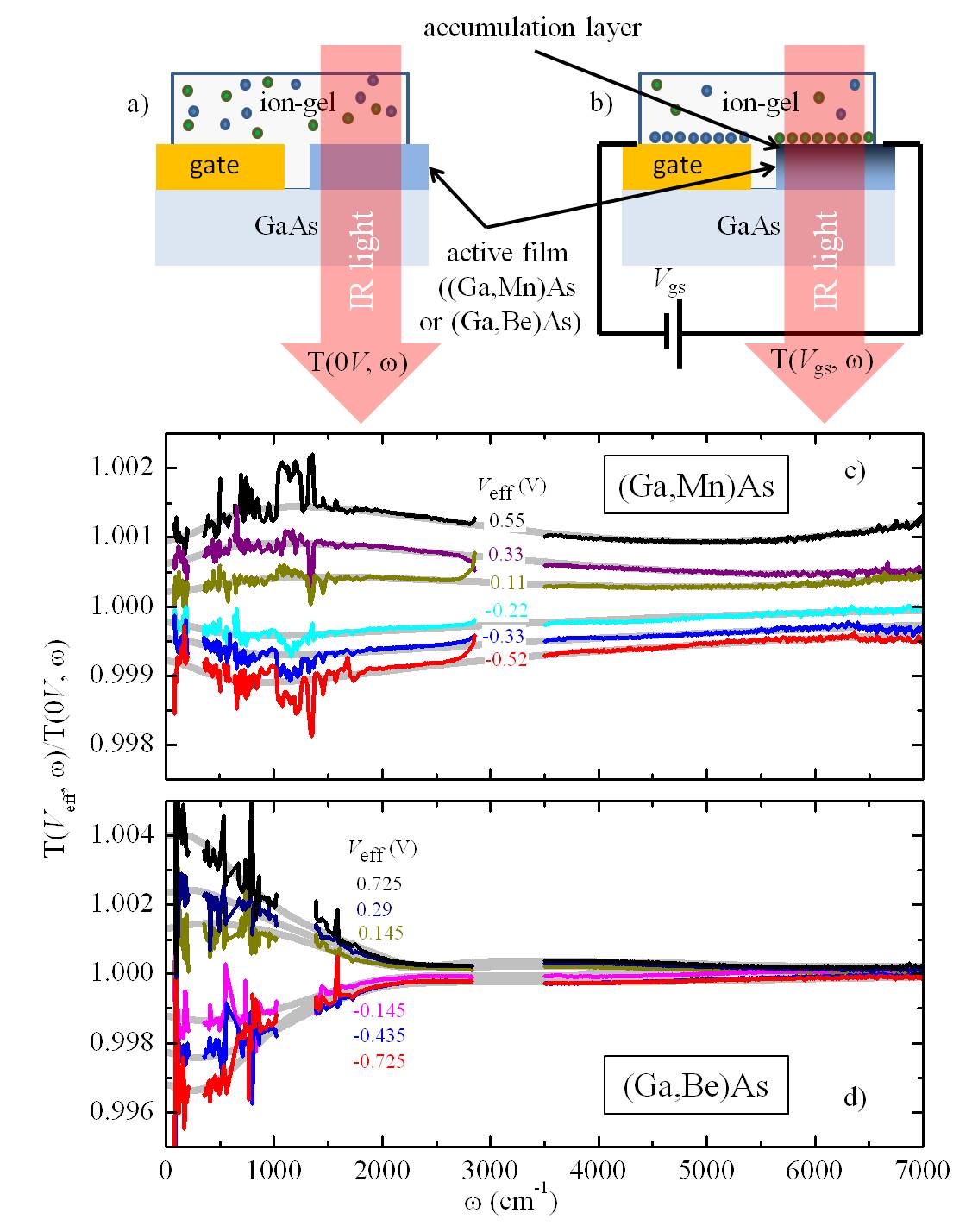}
\caption{Panels a and b highlight the operating principle of the EDL field-effect devices used in our IR experiments, as described in the text. Panel a shows a device in absence of any applied voltage, while panel b shows the device under applied $V_{gs}$. Panels c and d show the transmission spectra, under applied $V_{gs}$, through the $x$=0.015 Ga$_{1-x}$Mn$_x$As and $x$=3.2$\times$10$^{-4}$ Ga$_{1-x}$Be$_x$As based devices, normalized to their respective transmission spectrum at $V_{gs}$=0. The data are cut near 300 cm$^{-1}$ due to a GaAs phonon that eliminates transmission over the cut frequency range. The data are also cut where sharp features from strong vibrational modes in the ion-gel obscure the spectra (near 1200 cm$^{-1}$and 3200 cm$^{-1}$). Only select spectra are shown for clarity. The light gray lines in panels c and d are the model fits to the transmission spectra described in the text.
}
\label{schematic}
\end{figure}

In this work, we apply a gate-to-source voltage ($V_{gs}$) to produce an electric field-effect in electric double-layer (EDL) devices. The electric field-induced changes to the frequency dependent conductivity spectrum are probed by IR spectroscopy (see Fig.~\ref{schematic}). The active material in our EDL devices is either Ga$_{1-x}$Mn$_{x}$As, considered to be a prototypical ferromagnetic (FM) semiconductor, or a nonmagnetic $p$-doped counterpart system, Ga$_{1-x}$Be$_{x}$As. Both Be and Mn act as single acceptors in a GaAs host, however, Mn also adds a local magnetic moment. Therefore, Ga$_{1-x}$Be$_{x}$As serves as a useful and less complicated material to contrast with the magnetic Ga$_{1-x}$Mn$_{x}$As system. 

The advantages provided by field-effect studies are of extreme relevance to the physics of FM semiconductors in general, and of Ga$_{1-x}$Mn$_{x}$As in particular. The general agreement, and part of the technological appeal of Ga$_{1-x}$Mn$_{x}$As, is that the ferromagnetism is mediated by itinerant holes~\cite{Macdonald2005}. This fact was established through more than a decade of systematic exploration of Ga$_{1-x}$Mn$_{x}$As~\cite{Jungwirth2006, Burch2008, Sato2010, Dietl2010}, and validated through transport studies of the field-effect in Ga$_{1-x}$Mn$_{x}$As-based gated structures~\cite{Chiba2006, Sawicki2009, Endo2010}. Thus all proposed mechanisms of ferromagnetism in Ga$_{1-x}$Mn$_{x}$As are intimately tied to the dynamics of the charge carriers~\cite{Sato2010}. However, even after a decade of research, the details of the electronic structure and magnetic interactions remain in dispute~\cite{Dietl2010, Samarth2010, Samarth2012a}. A central open issue has been, and remains, whether the mediating holes reside in a disordered valence band (VB)~\cite{Dietl2000, Jungwirth2006} or in an impurity band (IB), which can be detached from the host VB or otherwise retains the $d$ character of the Mn dopants~\cite{Sato2003, Sato2010, Berciu2001, Mahadevan2004}.

The difficulty in obtaining a comprehensive understanding of the electronic structure of Ga$_{1-x}$Mn$_{x}$As is related to the high level of electronic disorder in the system. In order to synthesize Ga$_{1-x}$Mn$_{x}$As, low-temperature growth protocols are required~\cite{Ohno1998}. A direct but unintentional result of the low temperature growth, and a contributing factor to disorder in the system, are high concentrations of compensating interstitial and antisite defects~\cite{Edmonds2004, Missous1994}. The presence of these compensating defects assures the carrier density $p$ is never equal to the dopant density. However the exact relationship between the dopant concentration and $p$ cannot be established because of the myriad of complicated factors involved in the film growth. The uncertainty in carrier density and effects of disorder have clouded the interpretation of experiments~\cite{Jungwirth2007, Rokhinson2007, Richardella2010, Dobrowolska2012}, and proven an imposing theoretical challenge~\cite{Berciu2001, Yang2003, Moca2009, Kyrychenko2011}. Therefore, the systematic, disorder free addition or removal of carriers provided by the electric field-effect is highly desirable for studies of Ga$_{1-x}$Mn$_{x}$As. Moreover, the unique access to the electronic structure and dynamics granted by our IR probe establish IR spectroscopy as an ideal experimental tool to conduct such a study.

The films in the Ga$_{1-x}$Mn$_{x}$As-based devices were grown using low-temperature molecular beam epitaxy (MBE), on (001) GaAs substrates. This manuscript reports on a Ga$_{1-x}$Mn$_{x}$As-based device with active film grown to 100 nm, and a Mn doping density of $x$=0.015. The Mn doping density is inferred from reflection high energy electron diffraction intensity oscillation measurements of the MnAs growth rate~\cite{Schippan2000, Myers2006}. The doping level was chosen to keep $p$ as low as possible, yet still at a sufficient level to ensure the film is FM at low temperatures. Indeed, this (Ga,Mn)As film has a FM transition at $T_C$=25 K, determined by SQUID magnetometry. By limiting the carrier density in this way, the impact of the field-effect on our observables is maximized. We note, our results were reproduced on a second $x$=0.015 Ga$_{1-x}$Mn$_{x}$As-based device. The film of this latter divice was grown to 10 nm, under the same growth protocols as the 100 nm film. This latter device is referred to as Device B, and will be relevant to the intragap spectral weight comparison made later in the manuscript. The film in the Ga$_{1-x}$Be$_{x}$As-based device has a Be content of $x$=3.2$\times$10$^{-4}$, determined by Hall-effect measurements, placing it slightly above the critical IMT concentration ($x_c$=2.7$\times$10$^{-4}$) of Ga$_{1-x}$Be$_{x}$As~\cite{Nagai2005, Chapler2011}. The Be-doped film is 980 nm thick and was also grown on a (001) GaAs substrate. Ga$_{1-x}$Be$_{x}$As in this doping regime does not require low-temperature growth protocols. Thus the $x$=3.2$\times$10$^{-4}$ Ga$_{1-x}$Be$_{x}$As sample was grown using conventional equilibrium MBE. We note, IR spectroscopy data for the $x$=3.2$\times$10$^{-4}$ Ga$_{1-x}$Be$_{x}$As film was previously reported in Ref.~\cite{Chapler2011} by some of the same authors as this work. 

\begin{figure}
\centering
\includegraphics[width=86mm]{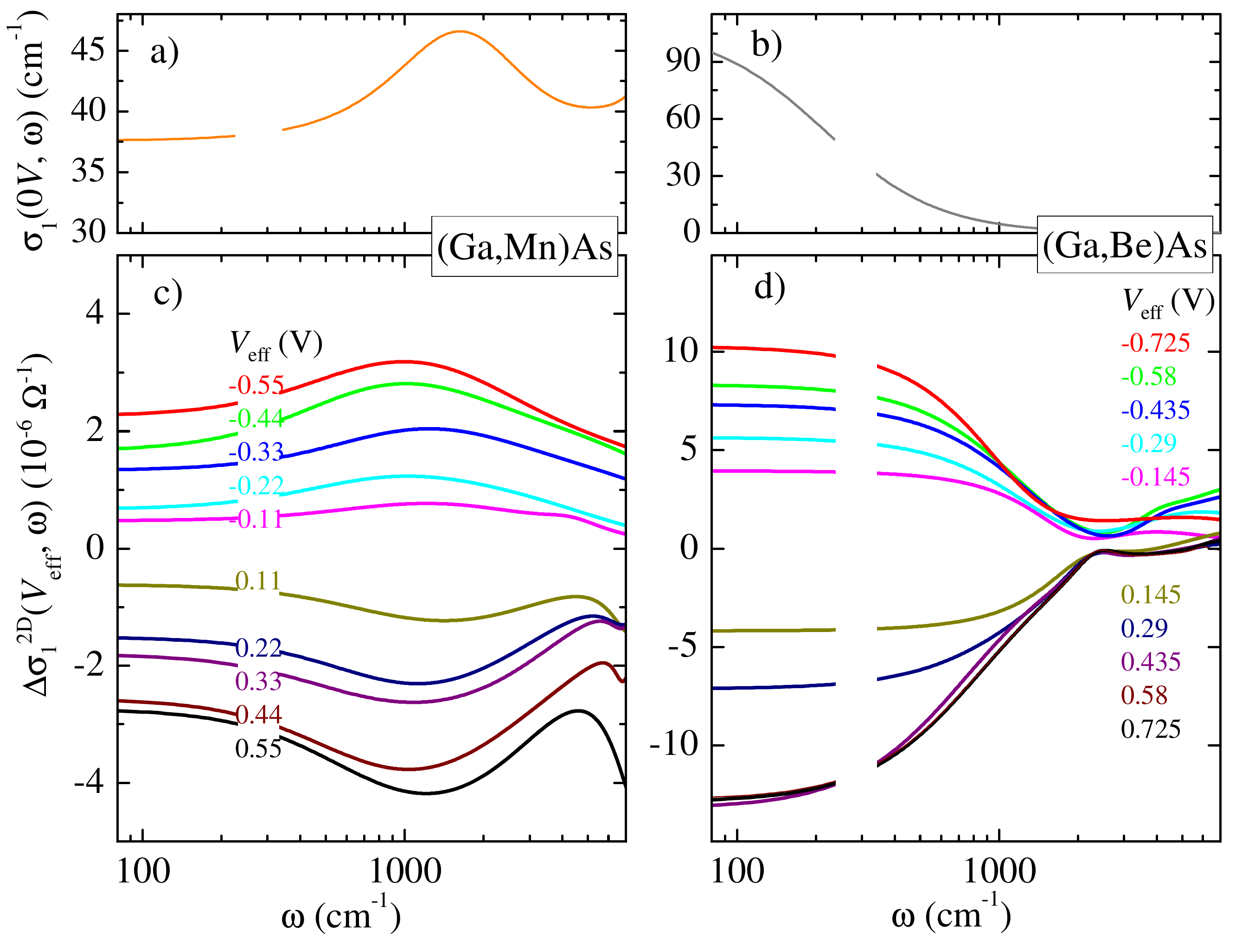}
\caption{Panel a and b show the $V_\mathrm{{eff}}$=0 IR conductivity spectrum of the $x$=0.015 Ga$_{1-x}$Mn$_x$As film and $x$=3.2$\times$10$^{-4}$ Ga$_{1-x}$Be$_x$As film used in our devices, respectively. The two-dimensional differential conductivity $\Delta$$\sigma^{2D}$($V_\mathrm{{eff}}$,$\omega$) spectra at all $V_\mathrm{{eff}}$ are shown for the (Ga,Mn)As-based device in panel c, and for the (Ga,Be)As-based device in panel d. The data are cut near 300 cm$^{-1}$ due to GaAs opacity over the cut frequency range.
}
\label{deltasigma}
\end{figure}

In order to quantify the effect of field-induced charges on the accumulation or depletion layer, it is imperative to extract the optical constants of the active film at $V_{gs}$=0.  These optical constants are necessary for the modeling of the transmission data of our EDL field-effect devices under applied $V_{gs}$ discussed later. To obtain the optical constants at $V_{gs}$=0, we first measure the transmission spectrum of the film on GaAs substrate, normalized by the bare substrate transmission. The transmission spectrum is then fit by a Kramers-Kronig (KK) consistent, multi-oscillator model. The model yields the complex conductivity spectrum of the film ($\sigma(0V, \omega)=\sigma_1(0V, \omega)+i\sigma_2(0V,\omega$)), provided the substrate optical constants are known or measured separately~\cite{Kuzmenko2009}. The real (or dissipative) part of the infrared conductivity spectra, $\sigma_1(0V, \omega)$, resulting from this analysis for the (Ga,Mn)As and (Ga,Be)As films can be seen in Figs.~\ref{deltasigma}a and b, respectively, and are discussed later in the manuscript.

Our EDL field-effect devices utilize an ``ion-gel'' (mixture of ionic liquid and block copolymer) to serve as the gate insulator. The ionic liquid used in our study is composed of 1-ethyl-3-methylimidazolium (the cation) and bis(trifluoromethylsulphonyl)imide (the anion). The block copolymer is polystyrene-poly(ethylene oxide)-polystyrene (PS-PEO-PS), and serves as a way to gel the ionic liquid for practical use in our devices. The composition of the ion-gel is approximately 75\% ionic liquid and 25\% block copolymer. The operating principle of EDL field-effect devices is that the mobile ions of the ionic liquid move to the device interfaces upon application of $V_{gs}$. This forms EDLs at the interfaces (see Fig.~\ref{schematic}b). The EDLs are effectively two capacitors in series, with one across the interface of the active film and ion-gel, and a second across that of the ion-gel and gate electrode. We thus define $V_\mathrm{{eff}}= \frac{A_g}{A_g+A_s}V_{gs}$, where $A_s$ is the gated area of the active film and $A_g$ that of the gate electrode, as the effective voltage drop across the interface of the active film and ion-gel. Advantages of ion-gels over conventional dielectrics include high charge injection density, and operation at low gate voltages~\cite{Lu2004, Shimotani2007, Dhoot2009, Das2008, Scherwitzl2009}. An additional advantage of ion-gels specific to IR studies is that devices can be designed such that the gate electrode is out of the optical path (see Fig.~\ref{schematic}a and b). If the gate electrode is $in$ the optical path, an IR transparent conductor must be used as the gate electrode. However, in this latter case, electrostatically induced effects in both the active film and the transparent conductor will be present in the IR spectra. We note, electrostatic effects in the transparent conductor observed in IR spectra can exceed those in the active film both in top and back gated devices~\cite{Li2007, Sai2007}.

In order to obtain the changes to the IR conductivity spectrum induced by application of $V_\mathrm{eff}$, transmission through the device is measured before, T($0V, \omega$), and after, T$(V_\mathrm{{eff}}, \omega$), $V_{gs}$ is applied. This procedure was repeated multiple times for each $V_{gs}$, and the transmission spectra were then averaged. The ratio T$(V_\mathrm{{eff}}, \omega$)/T($0V, \omega$) reveals the electrostatically induced changes to the spectrum, and is plotted in Fig.~\ref{schematic} for both types of device. It is assumed that neither the GaAs substrate nor the active film, apart from a narrow accumulation or depletion layer at the interface of the film and ion-gel, have any $V_{gs}$ dependence to their IR response. In the case of the ion-gel, electrostatically induced changes to the optical properties are confined to narrow vibrational modes, and do not show any broad effects. These assertions in regard to the ion-gel layer are confirmed by analysis of graphene-based devices~\cite{Ju2011}. We therefore attribute the broad changes to the spectra to the formation of a narrow accumulation or depletion layer at the interface of the active film and ion-gel.

To characterize the optical constants of the accumulation or depletion layer, T$(V_\mathrm{{eff}}, \omega$)/T($0V, \omega$) spectra were fit following the differential modeling protocol of the RefFit software package~\cite{Kuzmenko2009}. The differential modeling is done in two steps. The initial step is to build a base model. In our case, the base model describes the optical constants of the film at $V_{gs}$=0. This initial process of acquiring optical constants at $V_{gs}$=0 was described above. The second step, after the base modeling, is to construct a differential model, which represents the change in the optical constants of the accumulation or depletion layer. This latter model is comprised of multiple oscillators whose parameters are then adjusted in order to fit the change in the measured spectrum induced by some externally controlled parameter. The fits resulting from this modeling are shown as light gray lines in Figs.~\ref{schematic}c and d. The differential model yields the differential conductivity spectrum, $\Delta \sigma(V_\mathrm{{eff}}, \omega)=(\sigma(V_\mathrm{{eff}}, \omega)-\sigma(0V, \omega))$. This differential conductivity function satisfies the usual physical requirements of $\sigma(\omega)$, such as the KK relation.

We note the magnitude of both $\Delta \sigma(V_\mathrm{{eff}}, \omega)$ and $\sigma(V_\mathrm{{eff}}, \omega)$ extracted from modeling are dependent on the thickness $L$ of the accumulation or depletion layer used for the analysis.  Thus, in order to model the broad changes observed in the T$(V_\mathrm{{eff}}, \omega$)/T($0V, \omega$) spectra, $L$ must be assumed. We avoid this ambiguity by performing our analysis in two-dimensional units, and report the two-dimensional differential conductivity spectrum, $\Delta \sigma^{2D}(V_\mathrm{{eff}}, \omega)=\Delta \sigma(V_\mathrm{{eff}}, \omega)L$. The advantage of $\Delta \sigma^{2D}(V_\mathrm{{eff}},\omega$) is that the $L$ dependence is removed from the modeling, and thus $\Delta \sigma^{2D}(V_\mathrm{{eff}}, \omega)$ is independent of the actual thickness of the accumulation or depletion layer~\cite{Li2007, Tsui1978}. 

The $\Delta \sigma^{2D}_1(V_\mathrm{{eff}}, \omega)$ spectra, which represent changes to dissipative processes in the accumulation or depletion layer, of the (Ga,Be)As-based device are shown in Fig.~\ref{deltasigma}d. These data show monotonically increasing Drude-like lineforms (Drude lineform described by Eq.~\ref{Drude}) in the polarity associated with hole accumulation ($V_\mathrm{{eff}}<$0). In the reverse polarity ($V_\mathrm{{eff}}>$0), associated with hole depletion, monotonically decreasing inverted Drude-like lineforms are observed for $V_\mathrm{{eff}}$=0.0145 V to 0.435 V. At higher voltages, the effect appears to saturate, as the $V_\mathrm{{eff}}$=0.58 V and 0.725 V spectra are nearly identical to that of $V_\mathrm{{eff}}$=0.435 V.  

The $\sigma_1(0V, \omega)$ spectrum of the (Ga,Be)As film (Fig.~\ref{deltasigma}b) is described by a prominent Drude peak, with formula given by,

\begin{equation}
\sigma_1(\omega)=\frac{\Gamma \sigma_{dc}}{\Gamma^2+\omega^2}.
\label{Drude}
\end{equation}

\noindent The Drude lineform is characteristic of delocalized carriers in a metal. The amplitude at $\omega=0$ is equal to the dc conductivity $\sigma_{dc}$, and the width at half-max quantifies the scattering rate $\Gamma$. Thus, taking into account the Drude lineform of the $V_\mathrm{{eff}}$=0 spectrum, the $\Delta \sigma^{2D}_1(V_\mathrm{{eff}}, \omega)$ spectra can be understood as enhancing or diminishing the Drude response in the accumulation and depletion polarities respectively. The changes to the Drude response are extinguished by approximately 2500 cm$^{-1}$ at all $V_\mathrm{{eff}}$. Above 2500 cm$^{-1}$, much weaker changes to the spectra are observed, which are broad and featureless. As we show below in the sum-rule analysis of the (Ga,Be)As $\sigma_1(0V, \omega)$ spectrum, we conclude that the Fermi level ($E_F$) lies within the VB of the GaAs host. The conclusion of $E_F$ lying within the VB for this doping is also consistent with earlier IR studies of Ga$_{1-x}$Be$_{x}$As~\cite{Nagai2005, Chapler2011}. With $E_F$ established to be in the VB, we speculate that the changes above 2500 cm$^{-1}$ may be related to excitations from the split-off band~\cite{Braunstein1962} or broadening of the GaAs band gap edge. However, these effects are not examined in detail in this work. 

We note that although $E_F$ is established as residing in the GaAs host VB of the (Ga,Be)As film, there is no feature observed in any of the (Ga,Be)As spectra consistent with transitions from the light hole (LH) to heavy hole (HH) bands. Earlier IR experiments reported in Refs.~\cite{Nagai2005} and ~\cite{Chapler2011} have observed such a feature centered near 600-700 cm$^{-1}$ and 1800 cm$^{-1}$ in $x$=8.2$\times$10$^{-4}$ and $x$=0.009 Ga$_{1-x}$Be$_{x}$As, respectively. Based on these earlier data, any resonance due to LH$\rightarrow$HH transitions in the film reported here should be expected to be centered below 600 cm$^{-1}$, and weaker than the already relatively weak feature reported for the $x$=8.2$\times$10$^{-4}$ Ga$_{1-x}$Be$_{x}$As film of Ref.~\cite{Nagai2005}. Any electric field induced modifications to such a weak feature in this latter frequency regime, if present at all, may be obscured by the absorption lines of the ion-gel.

We now turn to the $\Delta \sigma^{2D}_1(V_\mathrm{{eff}}, \omega)$ spectra of the accumulation or depletion layer in the (Ga,Mn)As-based device, shown in Fig.~\ref{deltasigma}c. We first describe only the spectra in the hole accumulation polarity. In this polarity, the $\Delta \sigma^{2D}_1(V_\mathrm{{eff}}, \omega)$ spectra reveal a broad mid-IR resonance, peaked $\sim$1000 cm$^{-1}$, which monotonically increases in strength with voltage. Unfortunately, a more precise account of the peak frequency of the mid-IR resonance is unavailable. This limitation on the peak frequency is due to absorption modes in the ion gel, prevalent in the raw data, which obscure the spectra (see Fig~\ref{schematic}c). The $\Delta \sigma^{2D}_1(V_\mathrm{{eff}}, \omega)$ spectra also show enhancement at lower frequencies and in the limit of $\omega\rightarrow$0. At these far-IR frequencies, the spectra are relatively flat, and show monotonic increase in strength with voltage.

The $\sigma_1(0V, \omega)$ spectrum of the (Ga,Mn)As film (Fig.~\ref{deltasigma}a) shows a broad mid-IR resonance peaked near 2000 cm$^{-1}$, as well as relatively flat conductivity in the far-IR regime. These features are consistent with those reported in the literature for Ga$_{1-x}$Mn$_x$As~\cite{Singley2002, Burch2006, Jungwirth2010, Chapler2011}. The far-IR contribution to the $\sigma$$_1$(0$V$,$\omega$) spectrum is attributed to the electromagnetic response of itinerant carriers. Interpretation of the mid-IR resonance, however, remains a subject of debate~\cite{Burch2006, Jungwirth2007, Jungwirth2010, Chapler2011}. This latter feature lies in an energy range consistent with either transitions to an Mn-induced IB, where $E_F$ is within the IB; or intra-VB transitions from the LH to HH bands, with $E_F$ within the VB.

The far-IR and mid-IR features of the $V_\mathrm{{eff}}$=0 spectrum for the (Ga,Mn)As film suggest a natural interpretation of the $\Delta \sigma^{2D}_1(V_\mathrm{{eff}}, \omega)$ spectra. That is, we interpret $\Delta \sigma^{2D}_1(V_\mathrm{{eff}}, \omega)$ of the (Ga,Mn)As accumulation layer as enhancing both the itinerant carrier response and the strength of the mid-IR resonance. Although it is difficult to uniquely separate out these two contributions to $\Delta \sigma^{2D}_1(V_\mathrm{{eff}}, \omega)$, we note the T$(V_\mathrm{{eff}}, \omega$)/T($0V, \omega$) spectra cannot be reproduced without both contributions present in the model. In the depletion polarity, we observe similar features, however, with the lineshape ``inverted''. The strength of these depletion effects monotonically increase with voltage. Thus the features in the depletion polarity are interpreted as reducing the itinerant carrier response and strength of the mid-IR resonance of the (Ga,Mn)As film.

\begin{figure}
\centering
\includegraphics[width=64mm]{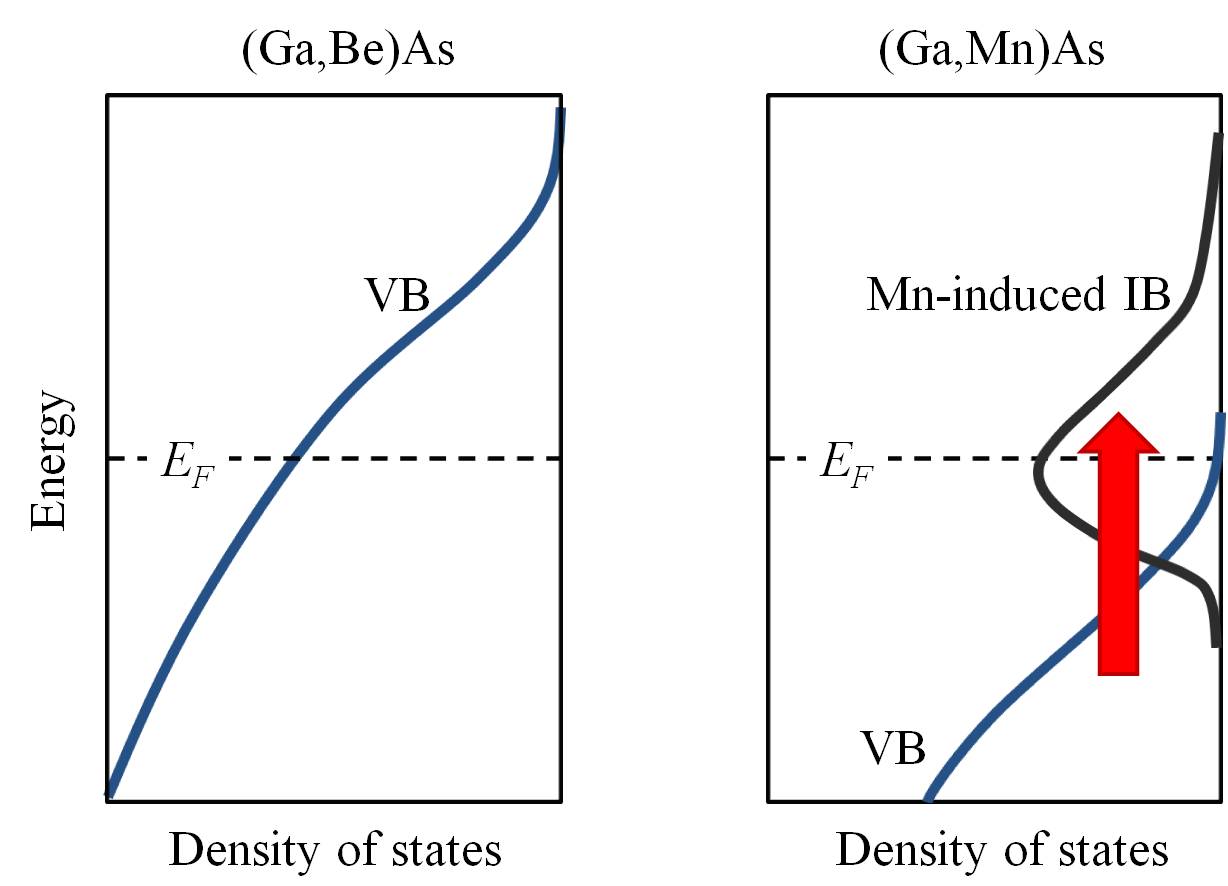}
\caption{Schematic illustrating the density of states in (Ga,Be)As (left panel) and (Ga,Mn)As (right panel) according to our experimental interpretation. The red arrow in the right panel represents transitions giving rise to the mid-IR resonance in the (Ga,Mn)As data. For these interband transitions, $m_\mathrm{{opt}}$ is approximately equal to the VB mass, as discussed in the text.
}
\label{bands}
\end{figure}

The contrasting forms of the spectral lineshapes of the (Ga,Be)As- and (Ga,Mn)As-based devices are consistent with detailed studies of the IMT in chemically doped Ga$_{1-x}$Be$_x$As and Ga$_{1-x}$Mn$_x$As~\cite{Singley2003, Burch2006, Chapler2011}. These latter studies show the signatures of the Mn-induced IB are present in the spectra of FM Ga$_{1-x}$Mn$_x$As films, and that the IB plays a vital role in the electrodynamics. Thus we attribute the broad mid-IR resonance of the (Ga,Mn)As-based device spectra to VB$\rightarrow$IB transitions. In contrast, the spectroscopic features of the (Ga,Be)As-based device data are all indicative of metallic conduction due to extended states within the host VB. These features consistent with VB conduction in the (Ga,Be)As-based device include the sum-rule analysis reported below. A schematic representation of the density of states consistent with our data is displayed in Fig.~\ref{bands}. 

In more common field-effect devices with oxide insulating material, the change in carrier density induced by the field-effect is determined by the dielectric constant, thickness of the oxide layer, and the applied voltage. Unfortunately in our devices, complexities of the EDLs prevent such simple calculations of the change in hole density induced by an electric field. However, a powerful aspect of IR spectroscopy is that it can serve as a contactless probe of the carrier density to effective mass ratio according to,
 
\begin{equation}
N_{\mathrm{eff}}(0V)=\frac{p(0V)}{m_\mathrm{{opt}}}=\frac{2}{\pi e^{2}}\int_{0}^{\omega_c}{\sigma_{1}(0V,\omega)d\omega},
\label{sum}
\end{equation} 

\noindent where $N_{\mathrm{eff}}$ is the ``spectral weight'', and $\omega_c$ is the appropriate cut-off for integration. When applied to a Drude peak alone, $m_\mathrm{{opt}}$ is equal to the effective band mass of the delocalized carriers~\cite{Basov2011}. In the (Ga,Be)As film, the hole density $p(0V)$ is known from room temperature Hall effect measurements to be 7.1$\times$10$^{18}$ cm$^{-3}$. Therefore we can extract $m_{\mathrm{opt}}$ of this film at $V_\mathrm{{eff}}$=0 by applying Eq.~\ref{sum} to the $\sigma_{1}(0V,\omega)$ spectrum in Fig.~\ref{deltasigma}b. From this analysis we find $m_{\mathrm{opt}}$=0.42 $m_e$ for our (Ga,Be)As film at $V_\mathrm{{eff}}$=0. This mass is in excellent agreement with the mass extracted from mobility data of $p$-type GaAs doped with nonmagnetic (Zn, C, or Be) acceptors~\cite{Alberi2008}. The experimental value of $m_{\mathrm{opt}}$ is in accord with conduction in the GaAs host VB, where the directionally averaged density-of-state masses of the HH and LH bands are $m_{HH}$=0.5 and $m_{LH}$=0.088, respectively~\cite{Blakemore1982}. Applying these HH and LH band masses to the expression for the two-band transport mass $m=(m_{HH}^{3/2}+m_{LH}^{3/2})/(m_{HH}^{1/2}+m_{LH}^{1/2})$, gives a mass of 0.38 $m_e$~\cite{Wiley1970}. 

Assuming no strong voltage dependence of $m_{\mathrm{opt}}$, we can relate the two-dimensional change in hole density $\Delta p^{2D}(V_\mathrm{{eff}})=(p(V_\mathrm{{eff}})-p(0V))L$ to the $\Delta \sigma^{2D}_{1}(V_\mathrm{{eff}},\omega)$ spectra by,

\begin{multline}
\Delta N^{2D}_{\mathrm{eff}}(V_\mathrm{{eff}})=\frac{\Delta p^{2D}(V_\mathrm{{eff}})}{m_\mathrm{{opt}}}\\ 
=\frac{2}{\pi e^{2}}\int_{0}^{\omega_c}{\Delta \sigma^{2D}_{1}(V_\mathrm{{eff}},\omega)d\omega}.
\label{deltasum}
\end{multline}

\begin{figure}
\centering
\includegraphics[width=86mm]{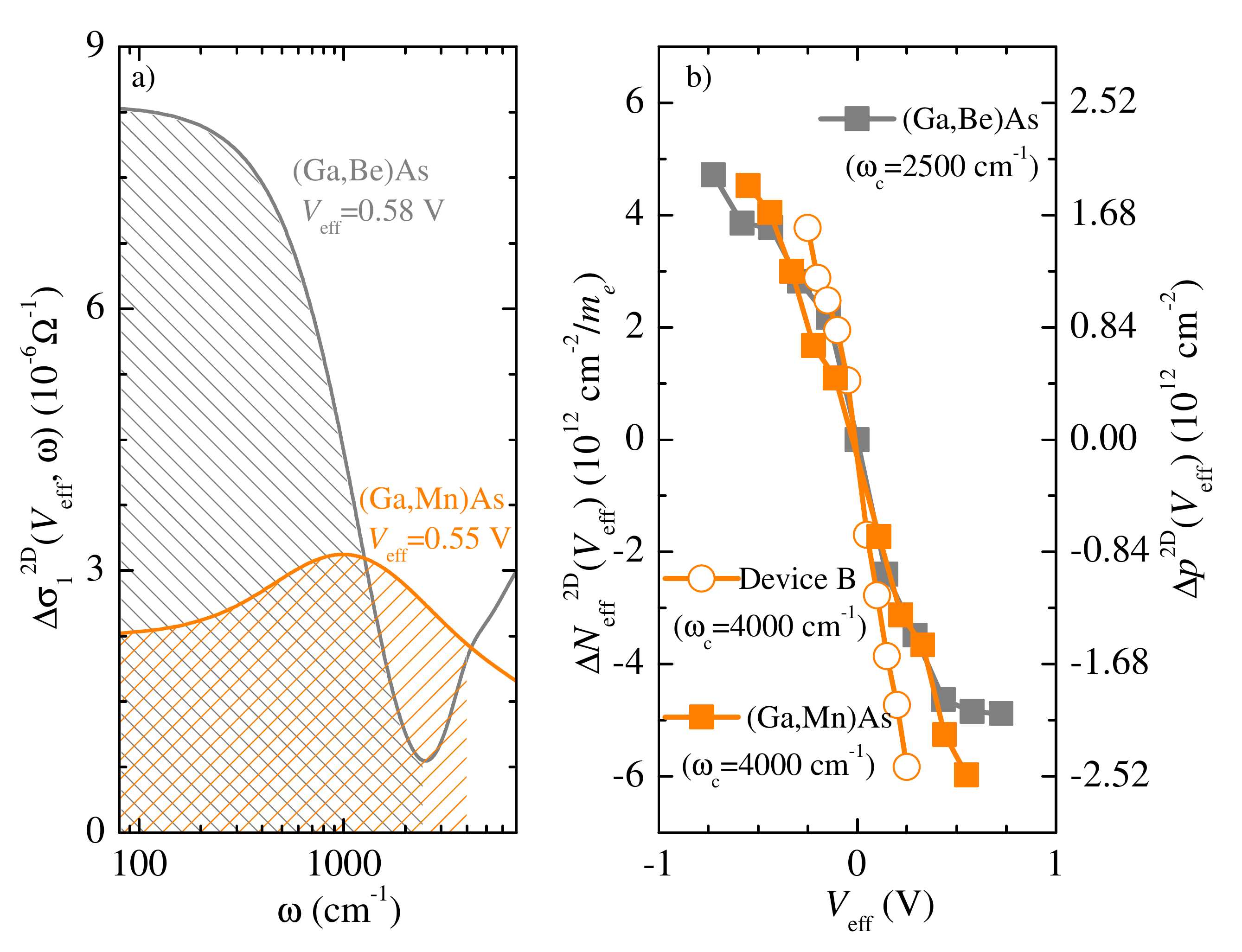}
\caption{Panel a shows the $\Delta \sigma^{2D}_1(V_\mathrm{{eff}}, \omega)$ for the (Ga,Mn)As-based device and (Ga,Be)As-based device at $V_\mathrm{{eff}}$=0.55 V and 0.58 V, respectively. The orange and gray shaded regions indicate the area included for equivalent intra-gap spectral weight in application of Eq.~\ref{deltasum} for the (Ga,Mn)As and (Ga,Be)As accumulation layers, respectfully. Panel b shows the change in spectral weight according to Eq.~\ref{deltasum} at all $V_\mathrm{{eff}}$ for the (Ga,Be)As, (Ga,Mn)As, and second (Ga,Mn)As (device B) based devices. The choice in integration cut-off $\omega_c$ is described in the text. Using the calibration procedure described in the text, the right axis is converted into the two-dimensional change in hole concentration $\Delta p^{2D}$.
}

\label{deltaN}
\end{figure}

\noindent For the (Ga,Be)As data of Fig.~\ref{deltasigma}d, we integrate the spectra according to Eq.~\ref{deltasum} up to $\omega_c$=2500 cm$^{-1}$. This cut-off serves to isolate only the portion of $\Delta \sigma^{2D}_1(V_\mathrm{{eff}}, \omega)$ attributed to modifications of the Drude response in the (Ga,Be)As-based accumulation or depletion layer. We plot $\Delta N^{2D}_{\mathrm{eff}}(V_\mathrm{{eff}}$) as a function of $V_\mathrm{{eff}}$ in Fig.~\ref{deltaN}. As $m_\mathrm{{opt}}$ has been independently determined for this film as described above, the right axis of Fig.~\ref{deltaN}b is scaled to reveal $\Delta p^{2D}(V_\mathrm{{eff}})$. In the hole accumulation polarity, $\Delta N^{2D}_{\mathrm{eff}}(V_\mathrm{{eff}}$) reveals sub-linear behavior with the effect equal to 4.7$\times$10$^{12}$ cm$^{-2}$/$m_e$ ($\Delta p^{2D}$=2.0$\times$10$^{12}$ cm$^{-2}$) at the largest voltage, $V_\mathrm{{eff}}$=-0.725 V. In the hole depletion polarity, $\Delta N^{2D}_{\mathrm{eff}}(V_\mathrm{{eff}}$) appears to saturate for $V_\mathrm{{eff}}>$0.4 V, as would be expected from the spectra. The largest effect in the hole depletion polarity is $\Delta N^{2D}_{\mathrm{eff}}(V_\mathrm{{eff}}$)=-4.9$\times$10$^{12}$ cm$^{-2}$/$m_e$ ($\Delta p^{2D}$=-2.1$\times$10$^{12}$ cm$^{-2}$) at $V_\mathrm{{eff}}$=0.725 V.

In the case of the (Ga,Mn)As data, a similar ``Drude-only'' sum-rule analysis is complicated due to the difficulty in uniquely separating the Drude component from the mid-IR resonance in the $\Delta \sigma^{2D}_1(V_\mathrm{{eff}}, \omega)$ spectra. Detailed quantitative analysis of estimates for the ``Drude mass'' in Ga$_{1-x}$Mn$_x$As can be found in Refs.~\cite{Singley2002, Burch2006}, which in general find it to be significantly larger than that of Ga$_{1-x}$Be$_{x}$As. Additional experimental probes and theoretical studies also support large carrier masses in Ga$_{1-x}$Mn$_x$As~\cite{Ohya2011, Burch2008, Alberi2008, Mayer2010, Bouzerar2011a, Moca2009a}. 

Another useful application of the sum-rule in Eq.~\ref{deltasum} is to examine the spectra over intragap energy scales (rather than ``Drude-only''). Thus, we extend the $\omega_c$ integration cut-off through the mid-IR resonance of the (Ga,Mn)As spectra, while holding it sufficiently low such that contributions to the spectral weight from excitations into the GaAs host conduction band are excluded. In Fig.~\ref{deltaN}a, the gray and orange shaded regions under the $\Delta \sigma^{2D}_1(V_\mathrm{{eff}}, \omega)$ spectra indicate the total area included in determining $\Delta N^{2D}_{\mathrm{eff}}(V_\mathrm{{eff}}$) of the $V_\mathrm{{eff}}$=0.58 V (Ga,Be)As ($\omega_c$=2500 cm$^{-1}$) and $V_\mathrm{{eff}}$=0.55 V (Ga,Mn)As ($\omega_c$=3800 cm$^{-1}$) accumulation layer spectra, respectively. The two shaded regions in the figure for these near equivalent voltages mark off an identical total area. In Fig.~\ref{deltaN}b, we generalize the intragap comparison of $\Delta N^{2D}_{\mathrm{eff}}(V_\mathrm{{eff}}$) of our materials. In this latter figure we show the change in intragap spectral weight of both devices at all $V_\mathrm{{eff}}$, with $\omega_c$=4000 cm$^{-1}$ and 2500 cm$^{-1}$ for the (Ga,Mn)As- and (Ga,Be)As-based devices, respectively. The data of Fig.~\ref{deltaN}a and b demonstrate that in general, the $\Delta N^{2D}_{\mathrm{eff}}(V_\mathrm{{eff}}$) found in the (Ga,Mn)As accumulation or depletion layer is roughly equivalent to that of the (Ga,Be)As accumulation or depletion layer for a given $V_\mathrm{{eff}}$.

Since the same ion-gel is used in both devices, the capacitance per unit area, and thus $\Delta p^{2D}(V_\mathrm{{eff}})$, are nominally similar in both (Ga,Mn)As- and (Ga,Be)As-based devices. We further justify this assumption by performing our IR experiments on a second $x$=0.015 Ga$_{1-x}$Mn$_x$As-based device. We show $\Delta N^{2D}_{\mathrm{eff}}(V_\mathrm{{eff}}$) with $\omega_c$=4000 cm$^{-1}$ for this device, labeled Device B, in Fig.~\ref{deltaN}b. We find a near identical trend in $\Delta N^{2D}_{\mathrm{eff}}(V_\mathrm{{eff}}$) for Device B as in the other two devices. The agreement of this trend between the two (Ga,Mn)As-based devices in particular, supports the conclusion that $\Delta p^{2D}(V_\mathrm{{eff}})$ is behaving nominally similar in all our devices. Moreover, the corresponding $\Delta N^{2D}_{\mathrm{eff}}(V_\mathrm{{eff}}$) per $\Delta p^{2D}(V_\mathrm{{eff}})$ suggests an approximate equivalence of $m_\mathrm{{opt}}$ as well.

We note, in the physical picture we have formulated from our data (see Fig.~\ref{bands}), the value of $m_\mathrm{{opt}}$ in the (Ga,Be)As- and (Ga,Mn)As-based devices is due to different processes. In the(Ga,Be)As-based device, $m_\mathrm{{opt}}$ is extracted from intraband (Drude) excitations within the VB. In contrast, $m_\mathrm{{opt}}$ of the (Ga,Mn)As-based devices is a result of integrating over the broad mid-IR resonance ascribed to interband processes (and relatively weak itinerant carrier response). For interband excitations, $m_\mathrm{{opt}}$ is the reduced mass of the bands involved~\cite{Dresselhaus}. Thus when applying the sum-rule over the mid-IR resonance (due to VB$\rightarrow$IB transitions) of the (Ga,Mn)As data, the reduced mass $m_\mathrm{{opt}}$ of the heavy IB and light VB will be approximately equal to the VB mass. The Drude mass of the (Ga,Be)As data is also clearly representative of the VB mass in that material, thus according to the physical picture of Fig.~\ref{bands}, an identical $m_\mathrm{{opt}}$ of our active materials implies an equivalence of their VB masses. The fact that the Drude mass of the (Ga,Be)As film is in accord with the VB mass of GaAs supports the conclusion that Mn or Be doping does not lead to a substantial renormalization of the GaAs host VB. This latter result is in agreement with resonant tunneling spectroscopy experiments on Ga$_{1-x}$Mn$_x$As layers~\cite{Ohya2011}.


Our results using this ``clean'' method of tuning the carrier density reveal the electrodynamics to be vastly different between the (Ga,Mn)As- and (Ga,Be)As-based devices. This difference indicates that conduction in FM Ga$_{1-x}$Mn$_x$As is distinct from genuine metallic behavior due to extended states in the host VB. The latter behavior is unmistakably exemplified by the clear enhancement or reduction of the Drude response in the accumulation or depletion layer of the (Ga,Be)As-based device. In contrast, the relatively weak effect of electrostatic doping on the far-IR conductivity, and the dominance of the broad mid-IR feature observed in the $\Delta \sigma_1(V_\mathrm{{eff}}, \omega)$ spectra of the (Ga,Mn)As-based device highlight the unconventional nature of conduction in this FM semiconductor. Moreover, the spectroscopic features observed in this work are all consistent with the ``IB picture'' of FM Ga$_{1-x}$Mn$_x$As.

A growing body of experimental evidence suggests the IB plays a central role in the electrodynamics of Ga$_{1-x}$Mn$_x$As~\cite{Okabayashi2001, Kojima2007, Ando2008, Mayer2010, Burch2008, Burch2006, Ohya2010a, Ohya2011, Chapler2011}. Due to the systematic nature of the addition or removal of carriers characteristic of the electric field-effect, this work provides a new level of detail on dynamical properties associated with the IB. We stress that theoretical description of the electronic structure and magnetism consistent with an IB model of Ga$_{1-x}$Mn$_x$As is also in place~\cite{Alberi2008, Bouzerar2011a, Bouzerar2010b, Berciu2001, Berciu2009, Moca2009a, Sato2003, Sato2010, Mahadevan2004, Tang2008}. This picture is in conflict, however, with the $p$-$d$ Zener model of ferromagnetism in Ga$_{1-x}$Mn$_{x}$As, where the FM interaction is mediated by holes in a weakly disordered VB~\cite{Dietl2000, Jungwirth2005}. On one hand, this latter model has successfully accounted for a number of observations related to the magnetic properties of Ga$_{1-x}$Mn$_x$As~\cite{Jungwirth2006, Sawicki2006, Sawicki2009, Nishitani2010}. On the other hand, there are also numerous experimental studies of the magnetic properties in dispute with the VB $p$-$d$ Zener model description of Ga$_{1-x}$Mn$_{x}$As~\cite{Rokhinson2007, Sheu2007, Mack2008, Dobrowolska2012}. Therefore, an alternative or competing FM mechanism in line with the IB picture, such as double exchange~\cite{Sheu2007, Sato2010}, should be thoroughly investigated and conclusively verified.



\begin{acknowledgements}
Work at UCSD is supported by the Office of Naval Research. Work at UCSB (Center for Spintronics and Quantum Computation) is supported by the Office of Naval Research and the National Science Foundation. Research at the UCSB (Center for Polymers and Organic Solids) is supported by the National Science Foundation DMR-0856060. Work at UC Berkeley (Department of Physics) is supported by the Department of Energy Early Career Award DE-SC0003949.
\end{acknowledgements}

\end{document}